\documentstyle[aas2pp4,epsf]{article}
\lefthead{W. L. Holzapfel \ea }
\righthead{Primary}

\slugcomment{To be submitted to {\em Astrophysical Journal}}

\def\D{{\Delta}}

\def\kl{{\rm{k}\lambda}}
\def\ea{et al.\ }
\def\ah{^{\rm h}}
\def\am{^{\rm m}}
\def\as{^{\rm s}}
\def\pr{^{\prime}}
\def\2pr{^{\prime \prime}}

\def\greatsim{\mathrel{\raise.3ex\hbox{$>$\kern-.75em\lower1ex\hbox{$\sim$}}}}
\def\lesssim{\mathrel{\raise.3ex\hbox{$<$\kern-.75em\lower1ex\hbox{$\sim$}}}}

\begin{document}

\title{Limits on Arcminute Scale Cosmic Microwave Background Anisotropy with 
the BIMA Array}

\author{
W.L.~Holzapfel\altaffilmark{1},
J.E.~Carlstrom\altaffilmark{2},
L.~Grego\altaffilmark{3},\\
G.~Holder\altaffilmark{2},
M.~Joy\altaffilmark{4},
and E.D.~Reese\altaffilmark{2}}


\altaffiltext{1}{Department of Physics, University of California,
Berkeley CA 94720}
\altaffiltext{2}{Department of Astronomy and Astrophysics, University of Chicago, Chicago
IL 60637}
\altaffiltext{3}{Harvard-Smithsonian Center for Astrophysics, Mail Stop 83,
60 Gardeen St., Cambridge MA 02138}
\altaffiltext{4}{Space Science Laboratory, SD50, NASA Marshall Space Flight Center,
Huntsville AL 35812}
\authoremail{swlh@cfpa.berkeley.edu}

\begin{abstract}
We have used the Berkeley-Illinois-Maryland-Association (BIMA) millimeter array
outfitted with sensitive cm-wave receivers to search for
Cosmic Microwave Background (CMB) anisotropies on arcminute scales.
The interferometer was placed in a compact
configuration which produces high brightness sensitivity, while
providing discrimination against point sources.
Operating at a frequency of $28.5\,$GHz, the {\it fwhm} primary beam of the
instrument is $\sim 6.6\pr$.
We have made sensitive images of seven fields, five of which where chosen 
specifically to have low IR dust contrast and be free of bright radio sources.
Additional observations with the Owens Valley Radio Observatory (OVRO)
millimeter array were used to assist in the location and removal
of radio point sources.
Applying a Bayesian analysis to the raw visibility data, we place
limits on CMB anisotropy flat-band power $Q_{flat}=5.6^{+3.0}_{-5.6}\,\mu$K 
and $Q_{flat}<14.1\,\mu$K at $68\%$ and $95\%$ confidence.
The sensitivity of this experiment to flat band power peaks at
a multipole of $\ell = 5470$ which corresponds to an angular scale of 
$\sim 2^{\pr}$. 
The most likely value of $Q_{flat}$ is similar to the level of the 
expected secondary anisotropies.
\end{abstract}


\section{Introduction} \label{sec:intro}
The Cosmic Microwave Background (CMB) has the potential to be a powerful 
probe of the early universe.
In the standard inflationary model, the CMB is imprinted with anisotropies that
reflect the distribution of matter at the epoch of recombination.
Observations of the CMB at degree angular scales probe structures which have
recently collapsed at this epoch and for which the distribution of anisotropies
is extremely sensitive to the cosmological model. 
On scales smaller than a few arcminutes, photon diffusion and the finite 
time of recombination damp the primordial fluctuations to near
zero amplitude (\cite{Hu97}).
However, the subsequent reionization of the universe can create a host of 
secondary anisotropies of the CMB.
For a review of the subject see \markcite{Haiman}Haiman \& Knox (1999).
On arcminute scales, secondary anisotropies generated since recombination are likely 
to dominate the primary signal.

There have been many previous searches for an-isotropy in the CMB on 
arcminute scales; both interferometric and single dish techniques 
have been used successfully.
In this paper we describe a search for arcminute-scale CMB anisotropies
with the Berkeley Illinois Maryland Association (BIMA) interferometer in a 
compact configuration at $28.5\,$GHz. 
We begin with a discussion of the instrument, field selection 
and observations in \S\ref{sec:instobs}.  
In \S\ref{sec:edit}, initial data reduction is described including point source 
detection and measurement.  
The Bayesian maximum likelihood analysis we apply to the data is described in 
\S\ref{sec:analysis}. 
The results of applying this formalism to our data are presented 
in \S\ref{sec:results} including a discussion of the effects of point source
subtraction. 
In \S\ref{sec:expsignals}, we discuss the levels of the expected signals.
We summarize the previous work in this field in \S\ref{sec:pwork}.
Finally, in \S\ref{sec:conclusion}, we summarize the results and discuss
prospects for future observations. 

\section{Instrument \& Observations} \label{sec:instobs}
The advent of low-noise, broad-band, millimeter-wave amplifiers has made
interferometry a particularly attractive technique for detecting and
imaging low contrast emission, such as anisotropy in the CMB.
An interferometer directly samples the Fourier transform of the intensity
distribution on the sky.
By transforming the interferometer output, images of the sky are obtained which
include angular scales determined by the size and spacing of the individual
array elements.
In this section, we describe the BIMA instrument, the selection of the
fields, and their observation.

\subsection{Instrument}
The anisotropy observations described here were made with nine elements 
of the BIMA array outfitted with 
sensitive cm-wavelength receivers.
The BIMA antennae are $6.1\,$m in diameter and produce $6.6\pr$ beams
at $28.5\,$GHz.
The receivers are based on low noise HEMT amplifiers
(\cite{Pospieszalski}) and are configured to respond only to 
right circularly polarized radiation.
At the operating frequency of $28.5\,$GHz,
the receiver noise temperatures range from $13-18\,$K and the
system temperatures were typically $35-55\,$K depending on source elevation.
The signals from the individual array elements are combined in the 
BIMA 2-bit digital correlator which was configured for these observations
with 8 contiguous $100\,$MHz sections of 32 channels each.
The array elements were placed in a compact 2-dimensional configuration
which provides high brightness sensitivity as well as sufficient resolution 
to identify radio point sources.

We have also made supporting observations with the Owens Valley Radio
Observatory (OVRO) array outfitted with the same cm-wave receivers.
These observations were used primarily for locating and measuring
point sources in the observed fields.
The OVRO antennae are $10.4\,$m in diameter and produce $3.8\pr$ beams
at $28.5\,$GHz.
For a description of the instrument see 
\markcite{Carlstrom96}Carlstrom \ea (1996). 
The BIMA and OVRO arrays, outfitted with cm-wavelength receivers, have 
been used to image more than 20 clusters (\cite{Carlstrom99}.

\subsection{Field Selection} \label{sec:fields}
In this paper, we report the results of a search for CMB anisotropy
in seven independent fields.
Two of the fields, PC1643+46 and VLA1312+32, are centered at previously
reported microwave decrements (\cite{Jones}; \cite{Richards97}). 
In a companion paper (\cite{noryle}), we demonstrate that our data are 
inconsistent with models used to describe the claimed decrements.
In 1997, we observed a third field in the direction of the quasar PSS0030+17
which was originally selected as a distant cluster candidate.
The quasar is at redshift $z=4.32$ and has two Ly$\,\alpha$ break
galaxies within $10\2pr$ (\cite{Djorgovski}). 
The way these three fields were selected would prevent us from claiming 
that they could produce an unbiased measurement of CMB anisotropy.
However, for the purpose of placing upper limits on CMB
anisotropy, we are justified in making use of these observations.

In 1998, four additional fields were selected to be evenly distributed in
right ascension and at convenient declinations for observations
with the BIMA array.
The fields were chosen to be in regions of low dust emission and contrast as 
determined from examination of IRAS $100\,\mu$m maps.
The VLA NVSS~(\cite{NVSS}) and FIRST~(\cite{FIRST}) surveys were then used to 
select regions free of bright point sources 
at $1.4\,$GHz.
Finally, we used the SkyView\footnote{We acknowledge the use of NASA's 
SkyView facility (http://skyview.gsfc.nasa.gov) located at NASA Goddard
Space Flight Center.} 
Digitized sky survey and ROSAT WFC maps
to check for bright optical or x-ray emission which could complicate 
follow-up observations.
The pointing centers for each of the 7 fields are given in 
Table~\ref{tab:coord}. 

\begin{table*}[t]
\begin{center}
\begin{tabular}{lcccc}
\multicolumn{5}{c}{Field Positions and Observation Times}\\\hline\hline 
\multicolumn{1}{c}{Fields} & $\alpha\,$(J2000) & $\delta\,$(J2000)  & Observation Year(s) & Time (Hours) \\ \hline
PC1643+46  &  $16\ah\,45\am\,11.3\as$ & $+46^{\circ}\,24\pr\,56\2pr$ & 1997, 1998 &$43.1$\\
VLA1312+32 &  $13\ah\,12\am\,17.4\as$ & $+42^{\circ}\,38\pr\,05\2pr$ & 1997 & $35.5$\\
PSS0030+17 &  $00\ah\,30\am\,16.4\as$ & $+17^{\circ}\,02\pr\,40\2pr$ & 1997 & $36.6$\\
BF0028+28  &  $00\ah\,28\am\,04.4\as$ & $+28^{\circ}\,23\pr\,06\2pr$ & 1998 & $77.6$\\
HDF1236+62 &  $12\ah\,36\am\,49.4\as$ & $+62^{\circ}\,12\pr\,58\2pr$ & 1998 & $32.0$\\
BF1821+59  &  $18\ah\,21\am\,00.0\as$ & $+59^{\circ}\,15\pr\,00\2pr$ & 1998 & $43.5$\\
BF0658+55  &  $06\ah\,58\am\,45.0\as$ & $+55^{\circ}\,17\pr\,00\2pr$ & 1998 & $44.7$\\
\end{tabular}
\end{center}
\caption{Coordinates of the observed fields, years of observation, and cumulative 
time on source.
} 
\label{tab:coord}
\end{table*}

For one of the new fields, we used the OVRO array to check for radio sources
at $28.5\,$GHz before beginning observations with the BIMA array.
With a 7 pointing mosaic, we reached a map {\it rms} flux of 
$\sim 120\,\mu$Jy  over a $8\pr$ region containing the entire BF0028+28 
field observed with BIMA.
We discovered a single source with a flux of $1.4\,$mJy.
This and all fluxes in this paper have been corrected for
the attenuation of the primary beam unless otherwise specified. 
The pointing center for the BIMA observations was chosen so that this 
source lied outside the observed field of view.
Unfortunately, we did not have time to image all the observed fields
with the OVRO array and it is possible that some of the fields observed
in 1998 suffer from low level point source contamination. 

\subsection{BIMA Observations}
All observations were made during the summers of 1997 and 1998,
interspersed between observations of the
Sunyaev-Zel'dovich effect (SZE) in x-ray selected clusters.
In 1998, we selected four fields spaced in right ascension so that
at any given time one of them had a hour angle suitable for observation.
Each $20$ minute source observation was bracketed by the observation of a 
calibration source.
Including the time for calibration cycles, the fraction of time spent 
on source was $\sim 60\%$. 
The integration times for each of the 7 fields are given in 
Table~\ref{tab:coord}. 
The fluxes of the calibration sources are all referenced to the flux 
of Mars which is uncertain by approximately $4\%$ (see discussion in 
\cite{Grego}).

\section{Editing and Inspection}    \label{sec:edit}
The data are edited using several criteria to ensure
the integrity of the calibration and that the results remain
free of systematics.
It is possible for the beam of one dish to be obscured by
one of its neighbors in the array. 
Baselines involving telescopes within $3\%$ of the shadowing limit are 
discarded. 
The spectral channels are inspected for any interference and
removed if they are believed to be contaminated.
Low signal to noise channels near the edges of the 
correlated bandwidth are not used. 
The effective noise bandwidth of the correlator after accounting
for the 2-bit digitization and removed end channels is $\sim 540\,$MHz.
Records with spurious system temperatures, caused by failed
or aborted calibration cycles are discarded. 
Source data not bracketed by successful calibration cycles are
discarded.
During periods of poor weather, the phase coherence of the calibration 
sources becomes poor.
All data that are bracketed by calibration cycles with poor phase 
coherence are also discarded.

\subsection{Point Sources} \label{sec:point}
Each data set is transformed to create a map with the DIFMAP
package (Shepard, Pearson, \& Taylor 1994). 
The maps are then searched for statistically significant unresolved emission.
In order to remain unbiased in our search for point sources, we limit ourselves to 
the range of baselines greater than $2.4\,\kl$  ($\ell>15000$),
which are completely independent of the baselines used in the anisotropy analysis.
That way, we can be assured that the anisotropy results will not 
systematically depend on the point source detection and subtraction.
In general, we find the flux and positions of the sources by fitting the Fourier 
transform of source model directly to the visibility data.
The source model is multiplied by the measured primary beam response to take 
the attenuation of the source into account.
In Table~\ref{tab:sources}, we list the positions and fluxes of the 
significant point sources in the observed fields.

\begin{table*}[t]
\begin{center}
\begin{tabular}{lcccccc}
\multicolumn{7}{c}{Subtracted Point Sources}\\\hline\hline 
\multicolumn{2}{c}{} &\multicolumn{2}{c}{Positions (J2000)} & \multicolumn{2}{c}{From Map Center} & $$ \\ 
\multicolumn{1}{c}{Field} & Source &  $\alpha$ & $\delta$  & $\Delta\alpha$ & $\Delta\delta$ & Flux ($\mu$Jy)\\ \hline
PSS0030+17 & $-$ &    $00\ah\,30\am\,37.0\as$ & $+17^{\circ}\,05\pr\,12\2pr$ & $294$ & $152$ & $12800\pm1600$ \\
HDF1236+62 & $-$ &    $12\ah\,36\am\,44.4\as$ & $+62^{\circ}\,11\pr\,33\2pr$ & $-34$ & $-85$ & $347\pm59$ \\
PC1643+46  & $S_1\pr$ &  $16\ah\,45\am\,21.1\as$ & $+46^{\circ}\,25\pr\,46\2pr$ & $102$ & $\phn 50$ & $227\pm70$ \\
PC1643+46  & $S_2\pr$ &  $16\ah\,45\am\,12.2\as$ & $+46^{\circ}\,24\pr\,08\2pr$ & $\phn \phn 9$ & $-48$ & $113\pm62$ \\
PC1643+46  & $S_3\pr$ &  $16\ah\,45\am\,22.7\as$ & $+46^{\circ}\,24\pr\,17\2pr$ & $118$ & $-40$ & $237\pm72$  \\
PC1643+46  & $S_1$ &  $16\ah\,45\am\,20.9\as$ & $+46^{\circ}\,25\pr\,43\2pr$ & $100$ & $\phn 47$ & $227\pm63$ \\
PC1643+46  & $S_2$ &  $16\ah\,45\am\,12.5\as$ & $+46^{\circ}\,24\pr\,13\2pr$ & $\phn \phn 6$ & $-43$ & $222\pm63$ \\
PC1643+46  & $S_3$ &  $16\ah\,45\am\,23.0\as$ & $+46^{\circ}\,24\pr\,21\2pr$ & $115$ & $-36$ & $262\pm71$  \\
\end{tabular}
\end{center}
\caption{Coordinates, distances from the anisotropy map center, and intrinsic fluxes 
of the significant sources in the field at $28.5\,$GHz.
For PC1643+46, we first show the results when the positions of the sources (primed) are 
fixed and second when the positions are allowed to assume their best fit values.
} 
\label{tab:sources}
\end{table*}

The brightest point source discovered was centered at   
$\alpha=00\ah\,30\am\,37.0\as$,  $\delta=+17^{\circ}\,05\pr\,12\2pr$ (J2000),
offset from the pointing center of the PSS0030+17 field by 
$\D \alpha=+294^{\2pr}$ and $\D \delta=+152^{\2pr}$.
The observed flux is attenuated by the finite size of the 
array element beams which are measured to have a {\it fwhm} 
of $\sim 396^{\2pr}$.
This source is far from the center of the map and has an observed flux
of $1.5\,$mJy.
Correcting for the primary beam response, the intrinsic source flux is 
determined to be $12.8\pm1.6\,$mJy.

A significant point source was detected in deep observations of the
HDF1236+62 field with both the OVRO and BIMA arrays. 
There were two pointings of the OVRO array with one centered on the 
position of the suspected point source.
We simultaneously fit the raw OVRO and BIMA visibility data with a single
component point source model.
The source is the brightest in the observed field with a flux of 
$345\pm59\,\mu$Jy.
It was centered at $\alpha=12\ah\,36\am\,44.4\as$,  
$\delta=+62^{\circ}\,11\pr\,33\2pr$ (J2000), offset from the BIMA pointing center by 
$\D \alpha=-34^{\2pr}$ and $\D \delta=-85^{\2pr}$.
This source has the same position as the brightest source found in a deep
radio image of the HDF at $8.4\,$GHz (\cite{Richards98}).

The field PC1643+46 was previously imaged with the Ryle telescope at $15\,$ GHz 
(\cite{Jones}).
Three point sources were found in the observed field.
In addition to our observations with the BIMA array, we have also imaged 
this same region with three pointings of the OVRO array operating at $28.5\,$GHz.
Each of the three pointings were chosen to place the map center near one of 
the suspected point sources. 
The field observed with BIMA was imaged with {\it rms} flux density 
ranging from $40-100\,\mu{\rm Jy}\,{\rm beam}^{-1}$.
We have performed a simultaneous fit to the BIMA and OVRO visibility data in 
order to determine the positions and fluxes of the sources at $28.5\,$GHz.
We have determined the fluxes of the sources using two different methods.
First we fixed the positions of the sources to the positions found with the 
Ryle observations and solved for the three source fluxes.
Introducing these three free parameters decreased the $\chi^2$ of the fit 
to the visibility data by 25; this indicates that the field suffers 
significant contamination from these sources.
The uncertainties for the source fluxes, shown in Table~\ref{tab:sources},
correspond to the change in flux which results in a increase in $\chi^2$ 
of one while the other free parameters (two other fluxes) are allowed to 
assume their best fit values.

We repeated this analysis allowing the source positions as well as fluxes to vary.
This procedure allows for differences in the positions determined by the BIMA and OVRO
analysis and those of \markcite{Jones}Jones \ea (1997). 
By allowing the positions to vary, and adding 6 new free parameters to the 
model, the $\chi^2$ of the fit is reduced by 14 from the case where the source positions
are fixed to the Ryle positions.
Therefore, the differences from the Ryle positions are significant and in the rest 
of this work we adopt these new source positions.
The differences between each of the new positions and those of 
\markcite{Jones}Jones \ea (1997) are less than $6\2pr$.
When the uncertainties in the Ryle positions and those found here 
are taken into account, the positions determined by the two experiments are found 
to be consistent.
The errors for each of the point source fluxes correspond to the change in flux 
required to produce to a change in $\chi^2$ of one while the free paramenters 
(flux of the other two sources and the positions 
of all three sources) are allowed to assume their best fit values.
In Table~\ref{tab:sources}, we list the measured positions and fluxes of
all the detected sources.
These sources are removed from the raw data by taking the Fourier transform
of the point source model modulated by the primary beam response and subtracting 
it directly from the visibility data.

\subsection{Image Statistics}
We have produced and analyzed images for each of the observed fields.
The results for the long baseline data used in point source subtraction 
are listed in Table~\ref{tab:lbimage}.
We limit the data to baselines $>2.4\,\kl$ to guarantee that the 
data used to determine the point source fluxes and positions are 
completely independent of the anisotropy data. 
The {\it rms} for all the data is considerably lower.  
The map {\it rms} indicates the accuracy with which the flux of point sources 
can be measured with this subset of the BIMA data. 
For the fields PC1643+46 and HDF1236+62 the point source sensitivity 
is considerably better than listed here due to the supporting OVRO observations.
The results using only the short baselines used 
in the anisotropy analysis are listed in Table~\ref{tab:sbimage}.
For the short baseline maps, we also express our results in terms of the
{\it rms} Rayleigh-Jeans (RJ) temperature fluctuations.
For both the short and long baseline maps, the observed {\it rms} values are 
compared to those expected from the noise properties of the visibilities.
For the short baseline results, there are approximately ten independent
beams in each observed field.
If we assume that the map values are dominated by Gaussian distributed noise,
the measured {\it rms} should be the same as the estimated value within  
approximately $10\%$ at $68\%$ confidence. 
 
\begin{table*}[htb]
\begin{center}
\begin{tabular}{lccc}
\multicolumn{4}{c}{Image Analysis for Baselines $>2.4\,\kl$}\\\hline\hline 
\multicolumn{1}{c}{Field} & beamsize($^{\2pr}$) & \multicolumn{2}{c}{{\it rms} ($\mu$Jy$\,$beam$^{-1}$)} \\ 
\multicolumn{2}{c}{} & estimated & measured\\ \hline
PC 1643+46  &  $18.9 \times 25.0$ & $132$ & $139$\\
VLA 1312+32 &  $18.8 \times 26.0$ & $134$ & $122$\\
PSS 0030+17 &  $18.1 \times 25.2$ & $155$ & $150$\\
BF 0028+28  &  $21.2 \times 21.6$ & $136$ & $127$\\
HDF 1236+62 &  $22.5 \times 23.9$ & $195$ & $207$\\
BF 1821+59  &  $21.7 \times 23.6$ & $152$ & $160$\\
BF 0658+55  &  $21.6 \times 23.2$ & $158$ & $157$\\
\end{tabular}
\end{center}
\caption{Image statistics for maps created using only the long baselines
used to measure point sources. Columns 3 \& 4 list the estimated and observed 
map {\it rms}.
} 
\label{tab:lbimage}
\end{table*}

\begin{table*}[htb]
\begin{center}
\begin{tabular}{lccccc}
\multicolumn{6}{c}{Image Analysis for $u$-$v$ Range $0.63-1.2\,\kl$}\\\hline\hline 
\multicolumn{1}{c}{Field} &  beamsize($^{\2pr}$) & \multicolumn{2}{c}{{\it rms} ($\mu$Jy$\,$beam$^{-1}$)} & \multicolumn{2}{c}{{\it rms} ($\mu$K)} \\ 
\multicolumn{2}{c}{} & estimated & measured & estimated & measured\\ \hline
PC 1643+46  &  $\phn 98.3 \times 116.1$  & $187$ & $191$ & $24.6$ & $25.1$\\
VLA 1312+32 &  $\phn 95.2 \times 113.4$  & $225$ & $223$ & $31.3$ & $31.0$\\
PSS 0030+17 &  $\phn 99.1 \times 115.5$  & $202$ & $203$ & $26.5$ & $26.6$\\
BF 0028+28  &  $108.9 \times 118.8$ & $127$ & $\phn 99$  & $14.7$ & $11.5$\\
HDF 1236+62 &  $110.9 \times 122.0$ & $206$ & $168$ & $22.8$ & $18.6$\\
BF 1821+59  &  $108.4 \times 122.4$ & $174$ & $224$ & $19.7$ & $22.4$\\
BF 0658+55  &  $105.2 \times 148.0$ & $252$ & $304$ & $24.3$ & $29.3$\\
\end{tabular}
\end{center}
\caption{Image statistics for maps created using only the short baselines 
used in the anisotropy analysis. Columns 5 \& 6 give the image results in 
{\it rms} RJ map temperature.
}
\label{tab:sbimage}
\end{table*}

\section{Analysis} \label{sec:analysis}
Several recent papers have dealt with the analysis of CMB data from
interferometers
(\cite{MarPar}; \cite{ATCA98}; \cite{HobLasJon}; \cite{HobMag}; \cite{Partridge};
\cite{White}).
In this work, we follow the formalism presented in 
\markcite{White}White \ea (1998) for the Bayesian analysis of CMB data.

In theories which predict Gaussian temperature fluctuations, the fundamental
theoretical construct is the correlation matrix of the measured data.  Since
the data are the visibilities measured at a set of points
${\bf u}_i$, we will need to know the correlation matrices for
the signal and noise of the observed visibilities.
The measured fluxes are given by
\begin{equation}
V({\bf u}) = {\partial B_\nu\over\partial T} \int d{\bf x}
  \ {\Delta T({\bf x})}\ A({\bf x})
  e^{2\pi i{\bf u}\cdot{\bf x}} \quad ,
\end{equation}
where $\D T(\bf x)$ is the temperature distribution on the sky, 
$A({\bf x})$ is the primary beam of the telescope, 
\begin{equation}
{\partial B_\nu\over\partial T} =
  2k_B \left( {k_BT\over hc} \right)^2 {x^4 e^x\over (e^x-1)^2} \quad ,
\end{equation}
$k_B$ is Boltzmann's constant, and $x\equiv h\nu/k_{B}T_{\rm cmb}$.
We define the visibility 
correlation matrix,  
\begin{eqnarray}
C_{ij}^{V} & \equiv & \left\langle V^*({\bf u}_i)V({\bf u}_j)\right\rangle \nonumber\\
	   &    =   & \left({\partial B_\nu\over\partial T}\right)^2
 \int_0^\infty w\,dw\; S(w) W_{ij}(w)\, ,
\end{eqnarray}
which is proportional to the product of the power spectrum, $S(w)$, 
and the visibility window function.
The window function is given by 
\begin{equation}
W_{ij}(|{\bf w}|) \equiv
    \int_0^{2\pi} d\theta_{w}\ \widetilde{A}^*({\bf u}_i-{\bf w})
    \widetilde{A}({\bf u}_j-{\bf w}) \quad , 
\label{eqn:wijdef}
\end{equation}
where ${\widetilde{A}({\bf u})}$ is the Fourier transform of the telescope
primary beam.
In the case of a single flat band power and $\ell > 60$, we can write 
\begin{equation}
C_{ij}^{V} = {6\over{5 \pi}} \left({\partial B_\nu}\over{\partial T} \right)^2
Q_{flat}^2 \int_0^\infty\frac{dw}{w} W_{ij}(w) \quad ,
\end{equation}
where
\begin{equation}
Q_{flat}^2 \equiv {5 \over {24 \pi}} C_{\ell}{\ell}{\left(\ell +1 \right)}
\end{equation}
is the normalization of the power spectrum.
The correlation function of the noise is diagonal with elements given by
\begin{equation}
C^N_{ii}={{1}\over{\sigma_i^2}}\quad ,
\end{equation}
where $\sigma_i$ is the variance of the measured visibilities.
For a given set of $n$ measured visibilities, one can test any theory, or
set of $\{C_\ell\}$, by constructing the likelihood function (for complex
visibilities)
\begin{equation}
{\cal L}\left( \{C_\ell\}\right) = {1\over\pi^n \det{C}} \ \exp
  \left[ -V^{*}({\bf u}_i) C_{ij}^{-1}V({\bf u}_j)\right] \quad,
\label{eqn:likelihood}
\end{equation}
where $C_{ij}=C^V_{ij}+C^N_{ij}$ is the correlation matrix of visibilities
at ${\bf u}_i$ and ${\bf u}_j$ (\cite{HobLasJon}).

\subsection{Joint Confidence Intervals} \label{sec:confidence}
Invoking Bayes' theorem and assuming a uniform prior for the amplitude of the 
fluctuations, we can determine the probability that the correct result is
contained in an interval {\it I},
\begin{equation}
P(I)={{\int_I{\cal L}(z) dz}\over{\int_0^{\infty}{\cal L}(z) dz}} \quad .
\end{equation}
The confidence interval corresponding to a probability $P_0$
is given by the $I_0$ such that
$P(I_0) = P_0$ and ${\cal L}[z \in I_0] \geq {\cal L}[z\not\in I_0]$.
If the fields are entirely independent, the joint likelihood
for the combination of the data sets to be described by a given model
is simply equal to the product of the likelihoods for the individual data sets,
\begin{equation} \label{eqn:jlike}
{\cal L}(z)=\prod_i{{\cal L}_{i}(z)}\quad .
\end{equation}
Combining the diagonal window functions corresponding to each visibility
weighted by the noise, we can construct an effective diagonal window 
function to determine where the experiment is most sensitive;
\begin{equation} \label{eqn:jwin}
{\bar W}_{\ell} = \sum_{i}{\frac{W_{ii}(\ell)w_i}{
\sum_{\ell}\frac{W_{ii}(\ell)}{\ell}\sum_{i}w_i}} \quad ,
\end{equation}
where $w_i=1/\sigma_i^2$.
With this normalization,
\begin{equation}
\sum_{\ell}{\frac{{\bar W}_{\ell}}{\ell}} = 1\quad .
\end{equation}
Using the data weighted window function, we can determine the effective multipole
of the experiment assuming the power spectrum is flat;
\begin{equation} \label{eqn:leff}
{\ell}_{eff} = \sum_{\ell} {\bar W}_{\ell} \quad . 
\end{equation}

\subsection{Binning} \label{sec:binning}
The data sets for each of our fields contain on the order of $N=10^5$
visibilities.
The inversion of the covariance matrix required for the analysis 
is a $N^{2.8}$ process (\cite{numrec}).
As discussed in \markcite{HobLasJones}Hobson, Lasenby \& Jones (1995), 
considerable compression of 
the data is necessary if the analysis is to be completed in a reasonable 
amount of computing time.
We divide the $u$-$v$ plane into a grid of cells; all the visibilities within
a cell are combined, weighted by the reciprocal of their estimated
noise variance;
\begin{equation}
V_{\alpha} ={{\sum\limits_{i \in \alpha}{{V_{i}}\over{\sigma_i^2}}}
\over {\sum\limits_{i \in \alpha}{1\over{\sigma_i^2}}}} \quad .
\end{equation}
We determine noise weighted $u$-$v$ positions just as we have 
determined the values for the visibilities;
\begin{equation}
u_{\alpha},v_{\alpha} ={{\sum\limits_{i \in \alpha}{{u_{i}}\over{\sigma_i^2}}}
\over {\sum\limits_{i \in \alpha}{1\over{\sigma_i^2}}}},
{{\sum\limits_{i \in \alpha}{{v_{i}}\over{\sigma_i^2}}}
\over {\sum\limits_{i \in \alpha}{1\over{\sigma_i^2}}}} \quad .
\end{equation}
The sampling theorem tells us that the Fourier transform of the sky is
completely specified by a sampling of the $u$-$v$ plane on a regular
grid with $\D u,\D v < 1/2 \theta_p$ where $\theta_p$ is the angular 
radius of the primary beam.
Because our final $u$-$v$ points do not form a regular grid, it is possible that 
we would have to decrease the size of a our grid by a factor of two to strictly 
satisfy this criterion.
Following \markcite{HobLasJones}Hobson, Lasenby \& Jones (1995), 
we define the extent of the beam as
the point at which the beam falls to $1\%$ of its peak value, 
$\theta_p=7.45\pr$.
We determine that the $u$-$v$ plane must be sampled more densely than
$\D u, \D v = 230$.
In practice, this is simple to achieve and we sample the $u$-$v$ plane at
intervals of $\D u, \D v = 60$ in the analysis presented in this paper.
This compresses the number of $u$-$v$ points to $\sim200$ for each of
the data sets.

\subsection{Calibration}
The likelihood analysis code has been checked by the analysis of 
simulated data sets. 
We took a data file from one of our observations, removed the visibilities, 
and replaced them with Gaussian distributed noise with the same weights as the 
original data. 
To this we added the Fourier transform of a realization of CMB anisotropy
with flat band power that had been modulated with the measured BIMA primary 
beam.
We produced 100 such data sets each with independent CMB and noise 
realizations.
The simulated data were then analyzed in exactly the same way as the 
real data treating each data set as an independent (uncorrelated) observation.
For an input flat-band power with $Q_{flat}=30\,\mu$K,
we found that the likelihood peaked at $Q_{flat}=31\,\mu$K with values 
$29-33\,\mu$K and $27-35\,\mu$K allowed at $68\%$ and $95\%$ confidence. 
We interpret this as a demonstration that the analysis code is correctly 
calibrated.

\section{Results} \label{sec:results}
In this section, we use the method that we have described above to
determine the relative likelihoods that the observed fields are described 
by a model for the CMB fluctuations with flat band power $Q_{flat}$. 

\begin{figure}[htb]
\plotone{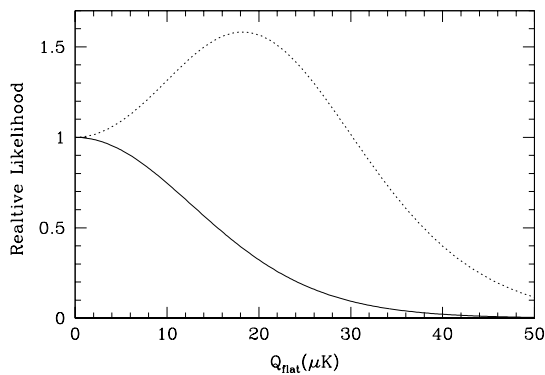}
\caption{The relative likelihood that the observed signal in the PC1643+46 field   
is described by flat band power with amplitude $Q_{flat}$.
The dotted line corresponds to an analysis ignoring the measured point sources and the
the solid line is the result when the measured point source fluxes 
are subtracted from the visibility data.
}
\label{fig:pntsub}
\end{figure}

Before proceeding, we discuss the effect of subtracting the known
point sources on the anisotropy results.
Three of the observed fields are known to have significant point source
contamination.
In Section~\ref{sec:point}, we discussed the determination of the source 
fluxes and positions.
For the three contaminated fields, we subtract the Fourier transform
of the point source model from the raw visibility data.  
In Figure~\ref{fig:pntsub}, we plot the results of the likelihood analysis 
for the field PC1643+46 before and after the subtraction of the three 
detected point sources.
The results are normalized to unity likelihood for the case of no 
anisotropy signal.
We have performed the same analysis with the point source fluxes found when 
the positions of the sources are fixed to the results of 
\markcite{Jones}Jones \ea (1997).
The results are essentially identical; the most likely value for $Q_{flat}$ 
is again zero.
In all further analysis, we use the point source fluxes and positions
found from the fits to the OVRO and BIMA data.
In Table~\ref{tab:pntsub}, we show the results for $Q_{flat}$ for
the analysis of the three contaminated fields before point source 
subtraction.
It is clear that  neglecting to subtract the known point sources
can lead to an erroneous detection of anisotropy.

\begin{table*}[htb]
\begin{center}
\begin{tabular}{lccc}
\multicolumn{4}{c}{Results Before Point Source Subtraction}\\\hline\hline 
$$ & \multicolumn{3}{c}{$Q_{flat}\,(\mu$K)}  \\ 
Field & Most Likely & $68\%$ & $95\%$ \\\hline
HDF 1236+62 & $\phn 4.8$ & $0.0-17.8$ & $0.0-33.2$ \\ 
PSS 0030+17 & $11.8$ & $0.0-31.6$ & $0.0-61.8$ \\ 
PC 1643+46 & $18.2$ & $4.8-29.0$ & $0.0-42.2$  \\ 
\end{tabular}
\end{center}
\caption{$Q_{flat}$ results before subtraction of known point sources.
Compare with Table~\ref{tab:qflat} to see the results after point source removal. 
}
\label{tab:pntsub}
\end{table*}

\begin{figure}[htb]
\plotone{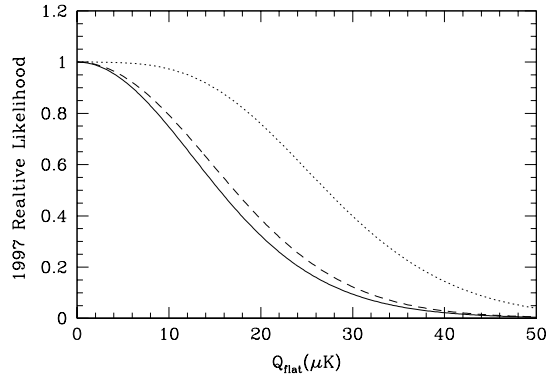}
\caption{The relative likelihood that the observed signal in each of the 
1997 data sets is described by flat band power with amplitude $Q_{flat}$.
The solid, dashed, and dotted lines are for the fields PC~1643+46, 
PSS~0030+17, and VLA~1312+32 respectively.
}
\label{fig:like97}
\end{figure}

\begin{figure}[htb]
\plotone{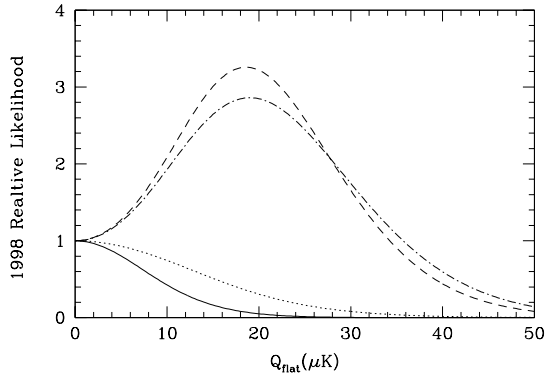}
\caption{The relative likelihood that the observed signal in each of the 
1998 data sets is described by flat band power with amplitude $Q_{flat}$.
The solid, dashed, dotted, and dot-dashed lines are for the BF~0028+28,
HDF~1236+62, BF~1821+59, and BF~0658+55 data respectively. 
}
\label{fig:like98}
\end{figure}

\begin{figure}[htb]
\plotone{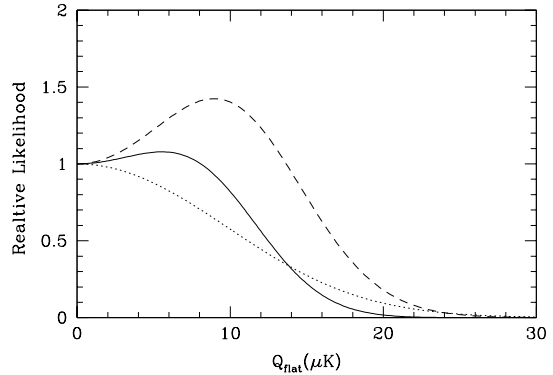}
\caption{The relative joint likelihood that the data from each year is described 
by flat band power with amplitude $Q_{flat}$.
The dotted, dashed, and solid lines correspond to the 1997, 1998, and 
combination of 1997 \& 1998 data respectively.
}
\label{fig:likeall}
\end{figure}

\begin{figure}[htb]
\plotone{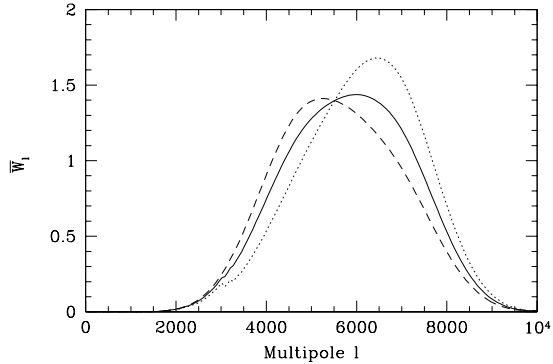}
\caption{The ``data weighted'' window functions $\bar{W}_{\ell}$. 
The dotted, dashed and solid lines correspond to the 1997, 1998,
and combination of 1997 \& 1998 data respectively. 
}
\label{fig:win}
\end{figure}

In Figures~\ref{fig:like97} and \ref{fig:like98}, we plot the relative 
likelihood for each of the observed fields, where we have subtracted
the detected point sources listed in Table~\ref{tab:sources}.
In Table~\ref{tab:qflat}, we list the $68\%$ and $95\%$ confidence intervals 
in $Q_{flat}$ for each of the observed fields. 
The fields are independent and we can apply eqn.~\ref{eqn:jlike}
to determine the joint likelihood for the combination of fields.
The relative likelihoods of the joint fits are plotted in 
Figure~\ref{fig:likeall}.  
In Table~\ref{tab:qflat}, we also list the confidence intervals in $Q_{flat}$ 
found from the joint fits.
Because of the different array configurations and declinations of the sources,
the window function for each observation is slightly different.
We have used eqn.~\ref{eqn:jwin} to determine effective diagonal window 
functions corresponding to the 1997, 1998, and combination of the 
1997 and 1998 data.
These window functions are plotted as a function of multipole in 
Figure~\ref{fig:win}.
Finally, we have used eqn.~\ref{eqn:leff} to determine the effective
multipole number, $\ell_{eff} \sim 5470$, of the results quoted here.

\begin{table*}[htb]
\begin{center}
\begin{tabular}{lccc}
\multicolumn{4}{c}{Most Likely $Q_{flat}$ and Confidence Intervals }\\\hline\hline 
\multicolumn{1}{c}{} & \multicolumn{3}{c}{$Q_{flat}(\mu{\rm K})$}\\ 
\multicolumn{1}{c}{Field} & Most likely & $68\%$ & $95\%$ \\ \hline
BF 0028+28   &   $\phn 0.0$ & $0.0-\phn 8.5$  &   $0.0-17.5$ \\
HDF 1236+62  &   $\phn 0.0$ & $0.0-13.0$ &   $0.0-26.6$ \\
BF 1821+59   &   $18.6$ & $8.8-28.8$ & $0.0-37.8$ \\
BF 0658+55   &   $19.0$ & $8.4-30.4$ & $0.0-40.6$ \\\hline
\multicolumn{1}{l}{Combined 1998 fields} & $\phn 8.8$ & $2.4-12.8$ & $0.0-17.4$\\
		&	&	&\\
PC 1643+46   &  $\phn 0.0$ & $0.0-13.1$ & $0.0-26.6$\\
VLA 1312+32  &  $\phn 0.0$ & $0.0-20.5$ & $0.0-38.0$\\
PSS 0030+17  &  $\phn 0.0$ & $0.0-14.4$ & $0.0-29.2$\\\hline
\multicolumn{1}{l}{Combined 1997 fields}  & $\phn 0.0$ & $0.0-\phn 8.9$ & $0.0-17.4$\\
\multicolumn{1}{l}{All fields} & $\phn 5.6$ & $0.0-\phn 9.6$ & $0.0-14.1$\\
\end{tabular}
\end{center}
\caption{Results of the Bayesian analysis for each of the observed fields.
Column 2 gives the most likely value for band power amplitude $Q_{flat}$.
Columns 3 \& 4 give the $68\%$ and $95\%$ confidence intervals.
}
\label{tab:qflat}
\end{table*}

One is immediately struck by the fact that the joint likelihood for all 
the data, shown in Figure~\ref{fig:likeall}, peaks at $Q_{flat}>0$.
However, this result has fairly low significance. 
Each of the individual fields are consistent with no signal on the sky at 
$95\%$ confidence and the confidence of a non-zero $Q_{flat}$ for the 
joint likelihood is only $44\%$.

After point source subtraction, none of the fields observed in 1997 
produce a non-zero anisotropy signal.
This is further evidence that the PC1643+46 and VLA1312+32 data are 
inconsistent with the presense of massive galaxy clusters in these 
fields as has been demonstrated by \markcite{noryle}Holzapfel \ea (1999).
There is also no evidence for a distant cluster in the PSS0030+17 field.
From Table~\ref{tab:qflat}, we can see that including the 1997 data in the 
joint likelihood only reduces the limits on $Q_{flat}$.

\subsection{Limits on Point Source Contamination} \label{sec:pntest}
All of the excess power is found in the two fields which have the 
poorest limits on point source contamination of any of the seven fields.
It is possible that point source contamination contributes to 
the signal in these fields.
If the observed power is dominated by Poisson distributed 
point sources, we expect to find $Q_{flat} \propto \ell$.
We have verified this scaling by the analysis of both real and 
simulated data with significant point source emission. 
If this scaling was observed in these data sets, it would be a 
``smoking gun'' for point source contamination. 

Two of the fields we have observed, BF0658+55 and BF1821+59, 
yield significant detections of power on short baselines. 
The flat-band power in these fields, determined from a joint 
analysis of the $0.63-1.2\,\kl$ data, is found to be 
$Q_{flat}=18.9\pm7.2\,\mu$K at $68\%$ confidence. 
To determine if this signal is due to point
source contamination, we have reanalyzed the data assuming   
$Q=Q_0(\ell/\ell_{0})$ rather than flat-band power.
The effective multipole $\ell_0$, to which $Q_0$ is referred, is chosen so 
that $Q_0=Q_{flat}$ when both analyses are applied to the 
$0.63-1.2\,\kl$ data.  
We have determined $Q_0$ for several fields with known point 
sources and find that,
for the long baselines ($1.2-6.0\,\kl$), the value of $Q_0$ is 
identical to that found from
the short baselines ($0.63-1.2\,\kl$) used in the anisotropy analysis.
Applying this analysis to the combined long baseline data for the fields 
BF0658+55 and BF1821+59, we find no evidence for point source 
emission and constrain $Q_0<8.8\,\mu$K and $Q_0<15.8\,\mu$K 
at $68\%$ and $95\%$ confidence.
Therefore, we conclude with $95\%$ confidence that 
Poisson distributed point sources cannot be responsible for  
excess power found in the analysis of the short baseline data.
However, if the signal is produced by several weak clustered point 
sources, such as in the PC1643+46 field, 
the scaling $Q_{flat} \propto \ell$ is not observed in either the 
real or simulated data.
Therefore, weak clustered sources could be responsible for the 
observed signal.

\section{Expected Signals} \label{sec:expsignals}
To interpret our results, it is informative to consider
the level of the expected signals. 
In this section, we give our best estimates of the expected 
contributions from primary and secondary CMB anisotropies and foreground
sources to the observed $Q_{flat}$. 
These results are compiled in Table~\ref{tab:signal}. 

\subsection{Primary Anisotropies}
We have convolved model CMB anisotropy power spectra with the window function 
of the experiment to determine the inferred flat band power signal,  
\begin{equation} \label{eqn:qflat}
Q_{flat}^2=\left({5}\over{24\pi}\right)\sum_{\ell}{C_{\ell} \ell \, {\bar W_l}} d\ell\quad .
\end{equation}
The CMB power spectra were generated using the CMBFAST code (\cite{Seljak}).
The expected signal for this experiment is largely determined by the total 
energy density of the universe, $\Omega_0$. 
We have calculated $Q_{flat}$ for a range of values for $\Omega_0$ while keeping 
the baryon density $\Omega_B=0.05$ and the Hubble constant $h=0.65$ constant.  
For these observations, the majority of the signal comes from the small
region of overlap between the BIMA array diagonal window function and the damping 
tail of the CMB. 
For total energy density $\Omega_0 = 1.0$ and $0.3$, we expect primary anisotropy 
signals of $Q_{flat}=1.1$ and $5.8\,\mu$K respectively. 

\subsection{Secondary Anisotropies}
Sometime after recombination at redshift $z \sim 1100$, the universe 
was reionized.
We know that this ionization was essentially complete by redshift $z\sim5$
because spectra of distant quasars do not show a continuum of absorption
by neutral hydrogen (\cite{Gunn}).
The interaction of the CMB with the reionized universe leads to secondary
anisotropies.
There are three types of secondary anisotropies which are expected to make 
significant contributions on arcminute scales:
the Vishniac effect; inhomogeneous reionization; and the 
SZ effect.

The Vishniac effect is a ``second order'' Doppler shift which produces 
CMB temperature anisotropy by the fact that the large scale velocity field is 
modulated by small-scale variations in baryon density (\cite{Vishniac};
\cite{Ostriker}).
While the other secondary anisotropies discussed in this section require 
variations 
in ionization, the Vishniac effect acts in a universe that is uniformly ionized.
\markcite{Hu96}Hu \& White (1996) have determined the size of the effect for a range
of reionization histories in the context of a critical CDM model.
For reionization epochs $z_r=5$ and $100$, they find signals of amplitude 
$Q_{flat} \sim 1.7\, {\rm and}\, 3.6\,\mu$K 
which peak at multipole moments $\ell \sim 5000$ and $10000$, respectively. 

Inhomogeneous reionization will imprint Doppler shifts, due to the velocities
of the reionized regions, on the Compton scattered CMB photons
(\cite{Kaiser}).
The effect of inhomogeneous reionization has been studied most recently by
\markcite{Gruzinov}Gruzinov \& Hu (1998) and 
\markcite{Knox}Knox, Scoccimarro, \& Dodelson (1998).
There are considerable uncertainties in the details of the generation of
ionized regions.
To obtain an accurate result, the correlation of the ionizing regions must 
be taken into account. 
For universes that reionizes at $z_i=26$ and $31$ , 
\markcite{Knox}Knox, Scoccimarro, \& Dodelson (1998) find a flat-band 
temperature anisotropy of $Q_{flat} \sim 1.8$ and $2.5\,\mu$K at $\ell=5500$. 
Observation of this signal would provide useful constraints
on what are presently highly speculative reionization scenarios. 

The majority of luminous matter in massive clusters of galaxies is observed to 
exist in the form of ionized gas which has been heated by gravitational infall.
This hot gas can present a considerable inverse Compton 
scattering cross section to CMB photons.
The resulting spectral distortion in the direction of a cluster
of galaxies is known as the Sunyaev-Zel'dovich effect (SZE) (\cite{SZ}).
The change in RJ CMB temperature in the direction of a 
massive cluster can be as large as $1\,$mK.
For a recent review see \markcite{Birkinshaw}Birkinshaw (1999) and the references 
within.
Several authors have computed the expected CMB anisotropy power 
spectrum due to the SZE in clusters of galaxies 
(\cite{Atrio}; \cite{Komatsu}; \cite{Holder}). 
The treatments differ in the range of cosmological models considered
and the models for the cluster evolution. 
One general conclusion is that, at the small angular scales relevant for this 
experiment, the majority of the signal 
is due to distant less massive clusters and removing either the bright
SZE or x-ray sources does not appreciably change the results. 
In general, the results depend sensitively on the assumed cosmology and 
cluster evolution model.
For example, \markcite{Holder}Holder \& Carlstrom (1999) find that at 
$\ell \sim 5000$, $Q_{flat} = 1.3-8.0\,\mu$K for the range of models they
consider. 

\subsection{Undetected Radio Point Sources} \label{sec:undetradio} 
In Section~\ref{sec:point}, we described our mechanism for measuring and removing
point sources.
As seen in Tables~\ref{tab:lbimage} and \ref{tab:sbimage}, 
the flux sensitivity of the data with baselines longer than $2.4\, \kl$ 
is comparable to that of the $0.63-1.2\,\kl$ data. 
Although we see no evidence for additional point sources, we cannot 
reliably constrain the presence of point sources below $3\sigma$ in the high resolution
maps which corresponds to a flux of $\sim 300-500\,\mu$Jy.

We have attempted to quantify the expected signals from point sources both 
analytically and through simulations.
The integrated source counts with fluxes less than $S_{cut}$ have been 
measured at $8.4\,$GHz by \markcite{Partridge}Partridge \ea (1997).
We take their result and scale it to the BIMA observing frequency of 
$28.5\,$GHz by using the 
average measured radio power law index ($\alpha=0.77$) from 
\markcite{Cooray98}Cooray \ea (1998).
The number density of sources then becomes
\begin{equation}
N(>S_{cut})={20 \over {\rm arcmin}^2} \left({\nu}\over{8.4\,{\rm GHz}}\right)^{-\beta \alpha} \left({S_{cut}}\over{\mu{\rm Jy}}\right)^{-\beta}\quad ,
\end{equation}
where $\beta=1.2$.  
We follow the treatment of \markcite{Scott}Scott \& White (1999) and estimate the 
contribution to the power spectrum to be
\begin{equation}
C_{\ell} \sim {{\beta}\over{\left({dB}\over{dT}\right)^2(2-\beta)}} N(>S_{cut}) S_{cut}^2 \quad ,
\end{equation}
where $S_{cut}$ is the minimum source flux which we can remove from our maps.
Using equation~\ref{eqn:qflat}, we can then determine the contribution of point
sources to $Q_{flat}$. 
For the maximum residual source flux, $S_{cut}=400\,\mu$Jy, we 
find $Q_{flat}=6.6\,\mu$K.

In order to test this approximation, we simulated distributions of point
sources on the sky.
For each simulated sky, we generated a sample of point sources with fluxes less 
than $S_{cut}=400\,\mu$Jy drawn from the $dN/dS$ distribution given in 
\markcite{Partridge}Partridge \ea (1997).
The sources are placed at random in the field.
These model skies are then Fourier transformed and added to a unique manifestation
of visibilities consistent with the weights of one of our complete data sets.
The simulations are then analyzed exactly as the real data. 
When we analyzed 100 manifestations of the sky generated in this way,
we found $Q_{flat}=4.8^{+2.0}_{-2.8}\,\mu$K at $68\%$ confidence.
So, the source simulations and analytic approximation predict similar
signals which, interestingly, are of the same order as the most likely signal 
in the data.

However, as described in Section~\ref{sec:intro}, the observed fields were selected
to be free of bright radio sources at $1.4\,$GHz. 
If the sources have a falling spectrum, we will have selected fields with 
significantly lower than typical point source confusion at $28.5\,$GHz.
Also, for the fields PC1643+46 and VLA1312+32, additional OVRO observations
were used to remove point sources down to $\sim 200\mu$Jy. 
Therefore, the estimates for point source contamination presented here are
upper limits to the expected signal in our data.

\subsection{Anomalous Dust Emission} \label{sec:dust}
Recently, anomalous foreground emission at microwave frequencies has been observed
which is found to be strongly correlated with IRAS $100\,\mu$m 
maps (\cite{Leitch}; \cite{deOliveira}).
It has been proposed that this emission may be due to either free-free emission 
(\cite{Kogut96})
or dipole emission from rapidly spinning dust grains (\cite{Draine}).
From a compilation of experimental results, \markcite{Kogut99}Kogut (1999) has 
determined a scaling between
emission in the IRAS $100\,\mu$m band and at microwave frequencies.
At $28.5\,$GHz, we expect this scaling to be approximately
$17\,\mu$K/(MJy$\,$sr$^{-1}$). 
As mentioned in Section~\ref{sec:fields}, we selected fields to have minimal 
$100\,\mu$m emission and contrast.
We have determined the {\it rms} $100\,\mu$m flux for each of our observed 
fields.
The resolution of the IRAS maps is $1.5\pr$ and therefore well matched to 
the angular scale on which the BIMA experiment is sensitive.
The observed fields are found to have a range of {\it rms} fluctuations 
$\D I_{100\mu} = 0.04-0.09\,{\rm MJy}\,{\rm sr}^{-1}$.
Therefore, we expect a {\it rms} temperature signal from this foreground
of $\D T < 1.7\,\mu$K, which corresponds to $Q_{flat}<1.1\,\mu$K.

\subsection{Systematic Errors}
The detected signals could also be the result of subtle systematic errors. 
The observations presented here represent the deepest images we have made of fields
without strong SZE decrements due to known galaxy clusters and therefore could be 
subject to undiscovered systematic errors.
Without success, we have extensively searched for a non-astronomical explanation of 
the observed excess power.
The results are found to be constant across the observing frequency band, independent 
of baseline or telescope, reproducible from day to day, and uncorrelated with the 
position of the sun or moon during our observations.
If this work is subject to systematic errors, deeper observations will be necessary 
in order for them to manifest themselves in a significant manner.

\begin{table}[htb]
\begin{center}
\begin{tabular}{lc}
\multicolumn{2}{c}{Expected Contributions to $Q_{flat}$}\\\hline\hline 
\multicolumn{1}{c}{Signal} & $Q_{flat}(\mu$K) \\\hline 
Primary Anisotropy 		    	& $1.1-5.8$\\
Vishniac Effect      			& $1.7-3.6$\\
Inhomogeneous Reionization  		& $1.8-2.5$\\
Sunyaev-Zel'dovich Effect 		& $1.3-8.0$\\
Point Sources 				& $< 6.6$\\
Spinning Dust/Free-Free 		& $< 1.1$\\\hline
Total					& $3.0-12.7$\\ 
\end{tabular}
\end{center}
\caption{Expected flat-band power due to primary anisotropy, secondary 
anisotropy, and foreground confusion.
}
\label{tab:signal}
\end{table}

\section{Comparison with Previous Work} \label{sec:pwork}
There here have been many previous searches for anisotropy in the CMB at
arcminute scales.
The results of this earlier work have been expressed in several different 
ways.
Until recently, it was common for experimenters to quote limits on CMB 
anisotropies with a Gaussian autocorrelation function (GACF)
\begin{equation}
C(\theta)=C_0\,exp{\left(-{{\theta^2}\over{2\theta_c^2}}\right)} \quad ,
\end{equation}
where $\theta_c$ is the coherence angle and $\sqrt{C_0}$ is the variance 
of the CMB.
Given the diagonal elements of the average window function, it is simple to 
convert between flat band power and GACF results (\cite{Bond95}).
Here we express our results in terms of limits on temperature anisotropy
with a GACF in order to facilitate comparison with the results of 
other experiments. 
At the scale of maximum sensitivity $\theta_c=0.9\pr$, our $68\%$ and 
$95\%$ confidence 
limits on $\sqrt{C_0}/T_{cmb}$ are $6.5 \times 10^{-6}$ and 
$9.6\times10^{-6}$ respectively.

In previous work, both single dish and interferometric techniques have 
been used to perform sensitive searches for CMB anisotropies on arcminute 
scales.
In Table~\ref{tab:pwork}, we list the frequency, sky coverage, coherence angle
corresponding to maximum sensitivity, 
and $95\%$ confidence limits on variance and flat-band power for each of the 
most sensitive experiments and compare them with our results.

\begin{table*}[htb]
\begin{center}
\begin{tabular}{lccccc}
\multicolumn{6}{c}{Comparison With Previous Work}\\\hline\hline 
\multicolumn{4}{c}{} & \multicolumn{2}{c}{$95\%$ Confidence Limits} \\ 
Experiment    & $\nu$ (GHz) & $\Omega_{sky}$ (arcmin$^2$) & $\theta_c$ (arcmin) & $\sqrt{C_0}/T_{cmb}$ & $Q_{flat}$ \\\hline
SuZIE         & $142\,$ & $\phn 213$     & $\phn \phn 1.1$ & $2.1 \times 10^{-5}$ & $-$\\
OVRO $40\,$m  & $20\,$  & $\sim 60$      & $\phn \phn 2.6$ & $1.7 \times 10^{-5}$ & $-$\\
VLA           & $8.4\,$ & $\phn \phn 20$ & $\sim 1.0$      & $-$                  & $35.2$\\
ATCA          & $8.4\,$ & $\phn \phn 28$ & $\phn \phn 1.0$ & $1.6 \times 10^{-5}$ & $23.6$\\
BIMA	      & $28.5$ & $\phn 240$ & $\phn \phn 0.9$ & $9.6 \times 10^{-6}$ & $14.1$\\

\end{tabular}
\end{center}
\caption{Frequency, sky coverage, coherence angle, and $95\%$ confidence limits 
on the variance and flat-band power from previous work and the BIMA results.
} 
\label{tab:pwork}
\end{table*}

Operating at a frequency of $20\,$GHz, the Owens Valley Radio Observatory 
(OVRO) $40\,$m dish has been used to measure sensitive differences
between beams of $\sim 1.8\pr$ {\it fwhm} separated by $\sim 7\pr$ (\cite{Readhead}).
They express their results in terms of limits on fluctuations with a GACF.
At the coherence angle for which the experiment is maximally sensitive,
$\theta_c=2.6\pr$, they constrain 
$\sqrt{C_0}/T_{cmb} < 1.7 \times 10^{-5}$ at $95\%$ confidence.
The effective total sky coverage of the experiment is estimated to be
$\sim 60\,{\rm arcmin}^2$.
More recently, the OVRO Ring experiment used the OVRO $40\,$m telescope to make 
a significant detection of anisotropy in a field near the North Celestial 
Pole (\cite{myers}).
These results are inconsistent with the earlier work at OVRO and are likely to 
be the result of foreground contamination.

The SuZIE experiment is a drift scanning bolometer array.
Fielded at the CalTech Submillimeter Observatory 
(CSO), it was used to map $\sim 213\,{\rm arcmin}^2$ of blank sky at 
its operating frequency of $142\,$GHz (\cite{Church}).
Unlike the other experiments discussed here, radio point sources are
not a significant source of confusion for SuZIE.
They also express their results in terms of limits on fluctuations
with a GACF.
At the coherence angle of maximum sensitivity, $\theta_c=1.1\pr$, 
they find $\sqrt{C_0}/T_{cmb} < 2.1 \times 10^{-5}$ at $95\%$ confidence.
 
Interferometers have also proved to be very effective for making sensitive
maps of the sky with arcminute resolution.
Using the Very Large Array (VLA) at $8.4\,$GHz, 
\markcite{Partridge}Partridge \ea (1997) obtained an extremely deep 
$21$ arcmin$^2$ image of the sky. 
On the scale at which the experiment is most sensitive 
(resolution $\sim 60\2pr$) they find $\D T/T_{cmb}< 2.0 \times 10^{-5}$ 
at $95\%$ confidence which corresponds to $Q_{flat} < 35.2\,\mu$K.
In order to achieve this limit, they are forced to subtract a statistical
estimate of the image variance due to point sources.
Because we do not have exact knowledge of the window function of the 
VLA observations, we cannot determine the response of their system to
CMB anisotropy with a GACF.

A group working with the Australian Telescope Compact Array (ATCA)
has recently produced what were, previous to this work, the lowest 
limits on arcminute scale CMB anisotropies (\cite{ATCA98}).
They observed at a lower frequency of $8.45\,$GHz, but the larger size
of the ATCA dishes compensates to make the window function of the ATCA 
and BIMA experiments similar. 
Using a single deep pointing of their array ($\sim 28\,$arcmin$^2$), they
constrained $Q_{flat}<23.6\,\mu$K at $95\%$ confidence on an angular scale 
corresponding to $\ell_{eff} \sim 4600$.
They also express their results in terms of anisotropy with GACF. 
At the coherence angle for which the experiment is 
maximally sensitive, $\theta_c=1.0\pr$, they find
$\sqrt{C_0}/T_{cmb} < 1.6 \times 10^{-5}$ at $95\%$ confidence.

\section{Conclusion} \label{sec:conclusion}
We have used the BIMA array in a compact configuration at 
$28.5\,$GHz to search for CMB anisotropy in seven independent fields.
With these observations, we have placed the lowest limits on 
arcminute scale CMB anisotropies to date.
These results are determined from $\sim 240\, {\rm arcmin}^2$ of sky; this 
is the largest sky coverage of any of the arcminute scale anisotropy experiments.
In the context of an assumed flat band power model for the CMB power spectrum, 
we find $Q_{flat}=5.6^{+3.0}_{-5.6}\,\mu$K at $68.3\%$ confidence and 
$Q_{flat}<14.1\,\mu$K at $95.4\%$ confidence with sensitivity centered about harmonic 
multipole $\ell_{eff} = 5470$.
This result includes the three fields observed in 1997 which 
were previously suspected to contain possible distant clusters.
None of these fields contribute to the observed excess power.

A detection of excess power is not surprising when one considers the level of 
the signals expected from secondary CMB anisotropies and foreground emission.
We have ruled out Poisson distributed point sources as the cause 
of the detected excess power at greater than $95\%$ confidence, although, faint 
clustered sources could still be responsible.
It is possible that we have detected some combination of secondary 
CMB anisotropies and faint radio point sources, however, the confidence 
of the detection is only $44\%$. 

In the coming year, we plan to expand our observations to include greater sky 
coverage and deep searches for point sources.
Future observations with broader correlated bandwidth could 
reach sensitivities  an order of magnitude higher than presented here. 
With proper characterization of foregrounds, these observations may be able to 
place interesting constraints on models for the reionization of the universe. 

Many thanks to the staff of the BIMA and OVRO observatories for their 
contributions to this project.
In particular, a high five to Dick Plambeck, Rick Forster, and John Lugten
for their help with the BIMA observations.
We would also like to thank Cheryl Alexander for her help in the construction of
the cm-wave receivers.
Thanks to Asantha Cooray and Sandy Patel for help with the OVRO and BIMA
observations. 
Radio Astronomy with the OVRO millimeter array is supported by NSF
grant AST 96-13717. 
The BIMA millimeter array is supported by NSF grant AST 96-13998.
JEC acknowledges support from a NSF-YI grant and the David and Lucile 
Packard Foundation.
EDR and LG acknowledge support from NASA GSRP fellowships.
This work is supported in part by NASA LTSA grant number NAG5-7986.
Finally, we would like to acknowledge informed discussions with Martin White
and Ravi Subrahmanyan.

\markright{REFERENCES}

\end{document}